# The Sequential Algorithm for Combined State of Charge and State of Health Estimation of Lithium Ion Battery based on Active Current Injection

Ziyou Song[a], Jun Hou[b], Xuefeng Li[c], Xiaogang Wu[c*], Xiaosong Hu[d*], Heath Hofmann[b], and Jing Sun[a]

[a] Department of Naval Architecture and Marine Engineering, University of Michigan, Ann Arbor, MI 48109, USA

[b] Department of Electrical Engineering and Computer Science, University of Michigan, Ann Arbor, MI 48109, USA

[c] College of Electrical and Electronics Engineering, Harbin University of Science and Technology, Harbin 150080, China

[d] Department of Automotive Engineering, Chongqing University, Chongqing 400044, China

**Abstract**—When State of Charge, State of Health, and parameters of the Lithium-ion battery are estimated simultaneously, the estimation accuracy is hard to be ensured due to uncertainties in the estimation process. To improve the estimation performance a sequential algorithm, which uses frequency scale separation and estimates parameters/states sequentially by injecting currents with different frequencies, is proposed in this paper. Specifically, by incorporating a high-pass filter, the parameters can be independently characterized by injecting high-frequency and medium-frequency currents, respectively. Using the estimated parameters, battery capacity and State of Charge can then be estimated concurrently. Experimental results show that the estimation accuracy of the proposed sequential algorithm is much better than the concurrent algorithm where all parameters/states are estimated simultaneously, and the computational cost can also be reduced. Finally, experiments are conducted under different temperatures to verify the effectiveness of the proposed algorithm for various battery capacities.

## 1. Introduction

Condition monitoring of lithium-ion batteries, such as the estimation of state of charge (SoC)

and state of health (SoH), is essential for practical applications [1]. Estimation of SoC and SoH is generally intertwined with estimation of battery parameters, which significantly vary with battery aging and changes in operating conditions [2] and are difficult to be adequately calibrated offline [3]. However, when all states/parameters are estimated simultaneously, substantial uncertainties are introduced in the estimation process, and the resulting inaccurate estimate of any parameter or state will dramatically impair the overall estimation performance. It has been proven that the estimation error of the multi-parameter estimation scenario (i.e., all states and parameters are estimated concurrently) is significantly increased when compared to the single-parameter estimation scenario (i.e., only one parameter or state is estimated) [4]. Moreover, multi-parameter estimation imposes a critical constraint on the battery current profile since a persistently exciting (PE) input condition should be satisfied to ensure convergence of the estimated parameters and states [5]. Generally speaking, the PE condition requires one frequency component for every two estimated parameters [6]. Therefore, it is worthwhile to investigate new algorithms which can separate the estimation of battery states from parameters, and therefore improve the estimation performance.

Model-based algorithms have been widely used for battery state estimation [7]. Commonly used models include the open circuit voltage (OCV) model [8], equivalent circuit model (ECM) [9], neural network model [10], and electrochemical model [11]. A previous study has shown that the first-order ECM is an acceptable choice for lithium-ion batteries due to its adequate fidelity and low computational cost [12]. The first-order ECM employed in this paper consists of an ohmic resistor $R_s$, an RC pair ($R_t$//$C_t$), and a DC voltage source $v_{oc}$, as shown in Fig. 1 [13]. In practical applications, these parameters, along with the SoC and SoH, should be estimated online [14].

Many algorithms have been developed and presented for SoC estimation. For example,

coulomb counting is a basic open-loop method [15], which however is dramatically influenced by inaccurate initial SoC and measurement noise [16]. The extended Kalman filter (EKF) is one of the most widely used methods [17]. In addition, unscented Kalman filters [18], sliding mode observers [19], and H∞ observers [20] have also been proposed in the literature. Similarly, for SoH, defined as the ratio of the remaining capacity to the original capacity [21], there are also many methodologies available for online implementation (e.g., adaptive EKF [22], multi-scale estimator [23], co-estimation method [24]). Moreover, combined SoC/SoH estimation has been specifically investigated using the standard dual-extended Kalman filter (DEKF) [25]. Even though the estimation algorithm is important, scholars have also focused on shaping the input-output data; i.e., the battery current and voltage, to further improve the estimation accuracy [26].

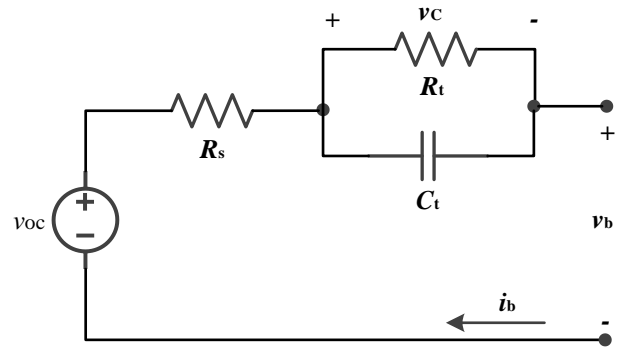

Fig. 1. The first-order equivalent circuit model for battery

An over-actuated system (e.g., hybrid energy storage system) offers an additional degree of freedom which provides the opportunity to inject desired input signals for identification objectives while simultaneously achieving control objectives [27]. To improve estimation performance, a sequential algorithm is proposed in this paper to estimate the parameters/states sequentially through active current injection. To exploit frequency-scale separation of different dynamics associated with different parameters and states the first-order ECM, which is governed by the initial SoC, the SoC variation, the ohmic resistance $R_s$, and the RC pair, is studied in simulation. It can be found that the initial SoC dynamics can be removed from the battery voltage dynamics with a high-pass filter. In addition, simulation results show that for the 18650 Lithium ion battery studied in this work, SoC variations can be neglected as long as the battery current frequency is not extremely low (i.e., below 0.001 Hz), and the ECM dynamic behavior is dominated by the resistance $R_s$ when the battery current frequency is high (i.e., above 0.1 Hz). Consequently, in Step #1, the proposed sequential algorithm estimates the ohmic resistance $R_s$ independently by injecting a high-frequency current and incorporating a high-pass filter, since the RC pair can be regarded as a short-circuit under these conditions. In Step #2, based on the estimated $R_s$, the RC pair can be characterized (i.e., the diffusion resistance $R_t$ and the time constant $\tau$) by injecting a medium-frequency current. Finally, in Step #3, the battery capacity and SoC can be estimated concurrently based on the above estimated parameters. The EKF is adopted in Steps #1 and #2 to estimate battery parameters, and a DEKF is adopted in Step #3 to estimate the battery SoC and SoH. The experimental results verify the effectiveness of the proposed sequential algorithm, which significantly increases the estimation accuracy when compared to the case where all parameters/states are estimated simultaneously.

## 2. System description

### 2.1 The first-order equivalent circuit model

Defining the battery terminal voltage as $v_b$ and the battery current as $i_b$ (positive for discharging and negative for charging), as shown in Fig. 1, the ECM dynamics are derived as follows:

$$\begin{cases} \dot{v}_C = -\frac{1}{C_t R_t} v_C + \frac{1}{C_t} i_b \\ v_b = v_{OC} - R_s i_b - v_C \end{cases}, \quad (1)$$

where $v_C$ is the RC pair voltage and $v_{OC}$ denotes the OCV. The OCV-SoC relation is given by [28]

$$v_{\text{OC}}(z) = K_0 - \frac{K_1}{z} - K_2 z + K_3 \ln(z) + K_4 \ln(1-z), \quad (2)$$

where $K_{0\text{-}4}$ are the constant coefficients and $z$ represents the normalized SoC, and the SoC dynamic is given as [29]

$$z = z_0 - \int_{t_0}^{t} \frac{\eta}{Q_\text{b}} i_\text{b}(t) \, \text{d}t, \quad (3)$$

where $z_0$, $\eta$, $t_0$, and $Q_\text{b}$ represent the initial SoC, the charging/discharging efficiency, the start time, and the battery capacity, respectively. To simplify the analysis, the OCV-SoC relationship is linearized as [29], [30]

$$v_{\text{OC}}(t) = a \left( z_0 - \int_{t_0}^{t} \frac{\eta i_\text{b}(\upsilon)}{Q_\text{b}} \, \text{d}\upsilon \right) + b, \quad (4)$$

where $a$ and $b$ are the corresponding coefficients. Note that this linearized relationship is only used in the analysis, while in the estimation process the nonlinear relationship shown in Eq. (2) is used. ECM parameters, including $R_\text{s}$, $R_\text{t}$, and $\tau$, are significantly influenced by battery degradation and operating condition, and are therefore difficult to calibrate for all practical scenarios [22]. Especially, the influence of degradation on the battery characteristics is almost impossible to be entirely investigated offline [31]. As a result, battery parameters should be estimated online along with battery SoC and SoH. Since battery parameters vary more slowly than the battery SoC [22], two assumptions are made in the estimation process:

1) The initial value of $v_\text{C}$ is 0, and

2) The parameters of $R_\text{s}$, $R_\text{t}$, and $\tau$ are assumed to be constant.

Under these assumptions, based on Eqs. (1) and (4), the transfer function from $i_\text{b}$ to $v_\text{b}$ is obtained.

$$v_\text{b}(s) = \frac{az_0}{s} + \frac{b}{s} - \frac{a}{s}\frac{\eta}{Q_\text{b}} i_\text{b}(s) - R_\text{s} i_\text{b}(s) - \frac{R_\text{t}}{1+\tau s} i_\text{b}(s), \quad (5)$$

where $s$ is the complex Laplace variable. There are therefore three parameters (i.e., $R_\text{s}$, $R_\text{t}$, and $C_\text{t}$)

and two states (i.e., $z_0$ and $Q_b$) in Eq. (5) that should be estimated.

## 2.2 The analysis of the first-order ECM dynamics

As shown in Eq. (5), the battery terminal voltage dynamics include four parts. The first part ($a \cdot z_0/s + b/s$) is constant and related to the initial SoC. The second ($-a \cdot \eta \cdot i_b(s)/s \cdot Q_b$) is related to the SoC variation and is significantly influenced by the battery capacity. The third ($-R_s \cdot i_b(s)$) is related to the ohmic resistance, and the fourth ($-R_t \cdot i_b(s)/(1+\tau \cdot s)$) is related to the RC pair. Since the first part is constant, it can be removed by a high-pass filter. A first-order high-pass filter is applied to Eq. (5), which yields:

$$v_{bf}(s) = \frac{(az_0+b)T_c}{1+T_c s} - \frac{a}{s}\frac{\eta}{Q_b} i_{bf}(s) - R_s i_{bf}(s) - \frac{R_t}{1+\tau s} i_{bf}(s), \quad (6)$$

where

$$\begin{cases} v_{bf}(s) = \dfrac{T_c s}{1+T_c s} v_b(s) \\ i_{bf}(s) = \dfrac{T_c s}{1+T_c s} i_b(s) \end{cases},$$

$T_c$ is the time coefficient of the high-pass filter, and $v_{bf}$ and $i_{bf}$ are the filtered battery voltage and current, respectively. The dynamics of the filtered system can then be presented in the time domain through the inverse Laplace transform. The effects of the initial SoC are given as

$$\mathcal{L}^{-1}\left\{\frac{(az_0+b)T_c}{1+T_c s}\right\} = (az_0+b)e^{-\frac{t}{T_c}}, \quad (7)$$

which will decay exponentially over time due to the high-pass filter, at the rate defined by $T_c$.

To evaluate the effects of the high-pass filter on separating the battery dynamics, we consider Samsung 18650 Lithium ion batteries. The parameters of an 18650 Lithium ion battery are specified in Table 1. The coefficients $K_{0\text{-}4}$ of the OCV-SoC for the adopted cell are 2.6031, 0.0674, -1.527, 0.6265, and -0.0297, respectively. The initial SoC dynamics vanish more quickly as $T_c$ decreases (i.e., the cut-off frequency increases). The initial SoC dynamics can be removed from the battery terminal voltage after ~6 minutes when $T_c$ is 80s. Therefore the initial SoC can be neglected in the filtered system and it will not influence the estimation of the other parameters/states.

**Table 1**

Specifications for the Samsung 18650 battery cell

| Parameter | Value |
|---|---|
| Nominal Voltage (V) | 3.63 |
| Cell Capacity (Ah) | 2.47 |
| Discharge/Charge Efficiency $\eta$ (%) | 98 |
| OCV-SoC slope $a$ (mV/100%) | ~8.845 |
| Standard Deviation of Voltage Measurement Noise $\sigma_V$ (mV) | 20 |
| Battery Current Amplitude $M$ (A) | 1 |
| Ohmic Resistance $R_s$ (mΩ) | ~100 |
| Diffusion Resistance $R_t$ (mΩ) | ~30 |
| Time Constant $\tau$ (s) | ~15 |

As shown in Eq. (6), except for the initial SoC response, the other responses are significantly influenced by the filtered current $i_{bf}$. To quantify the influence of the current's frequency on the voltage dynamics a sinusoidal current is considered in this paper:

$$\begin{cases} i_b(t) = M\cos(\omega t) \\ i_b(s) = \dfrac{Ms}{s^2 + \omega^2} \end{cases}, \quad (8)$$

where $M$ and $\omega$ are the current magnitude and frequency, respectively. To clearly show the influence of current frequency, the battery voltage responses when the current frequency is 0.4Hz, 0.004Hz, and 0.0004Hz are shown in Fig. 2, where $T_c$ is fixed at 80s. When the current frequency is 0.4Hz, as shown in Fig. 2 (a), the battery terminal voltage is governed by its ohmic resistance component, and the SoC variation component as well as the RC pair component can be neglected. The RC pair can be regarded as a short circuit when the current frequency is sufficiently high. In addition, the SoC variation component is small because high-frequency current does not induce a significant change in the battery SoC. This means that, based on the filtered signals, $R_s$ can be estimated independently by injecting high-frequency current regardless of the other parameters. As shown in Fig. 2 (b), the RC pair component is comparable with the ohmic resistance component when the current frequency decreases to 0.004Hz, while the SoC variation dynamics can still be neglected. Assuming that the parameters do not change quickly [22], the estimated $R_s$ can be used to estimate $R_t$ and $\tau$ when the current frequency is around 0.004Hz (i.e., medium-frequency current). As the current frequency further decreases to 0.0004Hz, as shown in Fig. 2 (c), none of three components can be neglected. The analysis presented above informs the sequential estimation algorithm presented in the next section.

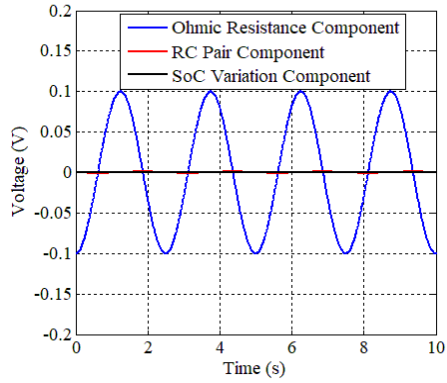

(a) $\omega$ is 0.4Hz

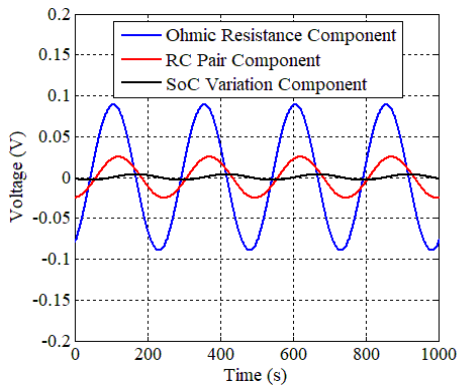

(b) $\omega$ is 0.004Hz

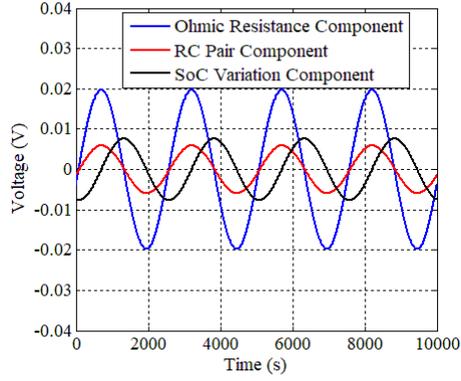

(c) $\omega$ is 0.0004Hz

Fig. 2. The battery voltage component due to various current frequencies

## 3. The sequential algorithm for combined SoC and SoH estimation

To reduce uncertainty in the estimation process and therefore increase the estimation accuracy, a sequential algorithm is proposed in this paper, as shown in Fig. 3. A high-pass filter is used to block the constant initial SoC component, and a high-frequency current is injected to estimate the ohmic resistance $R_s$. Using the estimated $R_s$, a medium-frequency current is injected to estimate the diffusion resistance $R_t$ and the time constant $\tau$. After obtaining the estimated $R_s$, $R_t$, and $\tau$, the SoC and SoH can be estimated finally.

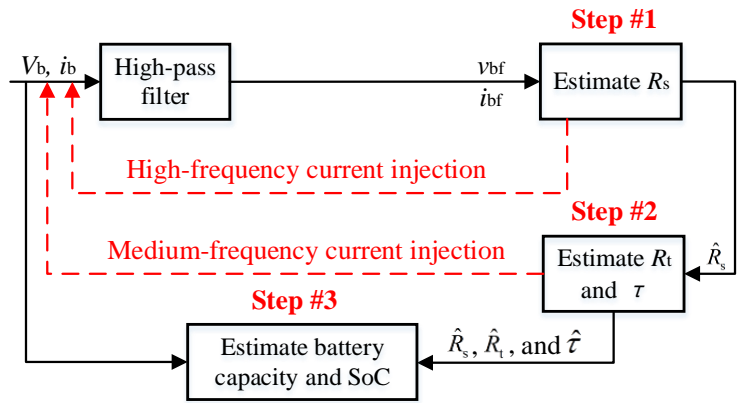

Fig. 3. The flowchart of the sequential algorithm

The sequential algorithm, which estimates the battery parameters and states separately, can reduce the computational cost when compared to the case when all states and parameters are estimated simultaneously [32]. More importantly, the sequential algorithm can improve the estimation accuracy since it exploits the frequency spectrum separation and eliminates the interactions of parameter and state estimation [26]. Three steps are involved in the sequential estimation, and the associated algorithms are presented herein.

### 3.1 A review of EKF and DEKF

The EKF is used in Steps #1 and #2 of the sequential algorithm to estimate the battery parameters. The EKF determines the optimal feedback gain to suppress both process noise and measurement noise [33]. The general discrete time state-space equation can be illustrated as

$$\begin{cases} \mathbf{\theta}_{k+1} = \mathbf{\theta}_k + \mathbf{r}_k \\ \mathbf{X}_{k+1} = \mathbf{H}(\mathbf{X}_k, \mathbf{\theta}_k, \mathbf{u}_k) + \mathbf{w}_k , \quad (9) \\ \mathbf{Y}_{k+1} = \mathbf{G}(\mathbf{X}_k, \mathbf{\theta}_k, \mathbf{u}_k) + \mathbf{v}_k \end{cases}$$

where $k$ is the time index, $\mathbf{X}_k$ is the state vector, $\mathbf{\theta}_k$ is the parameter vector, $\mathbf{u}_k$ is the input vector, $\mathbf{Y}_k$ is the output vector, $\mathbf{r}_k$ is the process noise for the parameters, $\mathbf{w}_k$ is the process noise for the states, and $\mathbf{v}_k$ is the measurement noise. In Steps #1 and #2, only the battery parameters are estimated, and the battery states are not involved. The calculation process of the EKF is summarized in Table 2. In Step #3, the combined SoC and SoH estimation is conducted based on the original system. We point out that the high-order OCV-SoC relationship (see Eq. (2)) is used to estimate SoC and SoH. The remaining battery capacity, which is also one of the battery parameters, is estimated to determine the SoH.

As a result, both the parameter (i.e., battery capacity) and the state (i.e., SoC) need to be estimated in Step #3. The DEKF method is a commonly used technique to simultaneously estimate states and parameters [34]. The DEKF adopts two EKFs run in parallel and estimates the state/parameter using each other's lastest updates [35]. Based on Eq. (9), the detailed algorithms of the DEKF is specified in Table 3.

**Table 2**

EKF algorithm

---

Initialization:

$$\begin{cases} \hat{\boldsymbol{\theta}}_0 = E[\boldsymbol{\theta}_0] \\ \boldsymbol{\Sigma}_{\boldsymbol{\theta}_0} = E\left[ (\boldsymbol{\theta}_0 - \hat{\boldsymbol{\theta}}_0)(\boldsymbol{\theta}_0 - \hat{\boldsymbol{\theta}}_0)^{\mathrm{T}} \right] \end{cases}, \quad (10)$$

where $\boldsymbol{\Sigma}_{\boldsymbol{\theta}_0}$ is the covariance matrix of parameter estimation error.

Parameter prediction:

$$\begin{cases} \hat{\boldsymbol{\theta}}_k^- = \hat{\boldsymbol{\theta}}_{k-1} \\ \boldsymbol{\Sigma}_{\boldsymbol{\theta}_k}^- = \boldsymbol{\Sigma}_{\boldsymbol{\theta}_{k-1}} + \boldsymbol{\Sigma}_{\mathbf{r}_{k-1}} \end{cases}, \quad (11)$$

where $\boldsymbol{\Sigma}_{\mathbf{r}_{k-1}}$ is the covariance matrix of process noise.

Parameter update:

$$\begin{cases} \mathbf{K}_k^{\boldsymbol{\theta}} = \boldsymbol{\Sigma}_{\boldsymbol{\theta}_{k-1}}^- \left(\mathbf{C}_{k-1}^{\boldsymbol{\theta}}\right)^{\mathrm{T}} \left[ \left(\mathbf{C}_{k-1}^{\boldsymbol{\theta}}\right) \boldsymbol{\Sigma}_{\boldsymbol{\theta}_{k-1}}^- \left(\mathbf{C}_{k-1}^{\boldsymbol{\theta}}\right)^{\mathrm{T}} + \boldsymbol{\Sigma}_{\mathbf{v}_{k-1}} \right] \\ \hat{\boldsymbol{\theta}}_k = \hat{\boldsymbol{\theta}}_k^- + \mathbf{K}_k^{\boldsymbol{\theta}} \left[ \mathbf{Y}_k - \mathbf{G}\left(\mathbf{X}_{k-1}, \hat{\boldsymbol{\theta}}_k^-, \mathbf{u}_k\right) \right] \\ \boldsymbol{\Sigma}_{\boldsymbol{\theta}_k} = \left(\mathbf{I} - \mathbf{K}_k^{\boldsymbol{\theta}} \mathbf{C}_{k-1}^{\boldsymbol{\theta}}\right) \boldsymbol{\Sigma}_{\boldsymbol{\theta}_{k-1}}^- \end{cases}, \quad (12)$$

where $\mathbf{C}_{k-1}^{\boldsymbol{\theta}} = \left.\dfrac{\partial \mathbf{G}(\mathbf{X}_{k-1}, \boldsymbol{\theta}, \mathbf{u}_k)}{\partial \boldsymbol{\theta}}\right|_{\boldsymbol{\theta}=\hat{\boldsymbol{\theta}}_k^-}$.

---

**Table 3**

DEKF algorithm

---

Initialization:

$$\begin{cases} \hat{\boldsymbol{\theta}}_0 = E[\boldsymbol{\theta}_0] \\ \boldsymbol{\Sigma}_{\boldsymbol{\theta}_0} = E\left[ (\boldsymbol{\theta}_0 - \hat{\boldsymbol{\theta}}_0)(\boldsymbol{\theta}_0 - \hat{\boldsymbol{\theta}}_0)^{\mathrm{T}} \right] \\ \hat{\mathbf{X}}_0 = E[\mathbf{X}_0] \\ \boldsymbol{\Sigma}_{\mathbf{X}_0} = E\left[ (\mathbf{X}_0 - \hat{\mathbf{X}}_0)(\mathbf{X}_0 - \hat{\mathbf{X}}_0)^{\mathrm{T}} \right] \end{cases}, \quad (13)$$

where $\boldsymbol{\Sigma}_{\mathbf{X}_0}$ is the covariance matrix of state estimation error.

Parameter prediction:

$$\begin{cases} \hat{\boldsymbol{\theta}}_k^- = \hat{\boldsymbol{\theta}}_{k-1} \\ \boldsymbol{\Sigma}_{\boldsymbol{\theta}_k}^- = \boldsymbol{\Sigma}_{\boldsymbol{\theta}_{k-1}} + \boldsymbol{\Sigma}_{\mathbf{r}_{k-1}} \end{cases}, \quad (14)$$

State prediction:

---

$$\begin{cases} \hat{\mathbf{X}}_k^- = \mathbf{H}\left(\hat{\mathbf{X}}_{k-1}, \hat{\boldsymbol{\theta}}_k^-, \mathbf{u}_k\right) \\ \Sigma_{\mathbf{X}_k}^- = \mathbf{A}_k \Sigma_{\mathbf{X}_{k-1}} \mathbf{A}_k^T + \Sigma_{\mathbf{w}_k} \end{cases}, \quad (15)$$

where $\mathbf{A}_k = \dfrac{\partial \mathbf{H}(\mathbf{X}, \hat{\boldsymbol{\theta}}_k^-, \mathbf{u}_k)}{\partial \mathbf{X}}\bigg|_{\mathbf{X}=\hat{\mathbf{X}}_k^-}$.

State update:

$$\begin{cases} \mathbf{K}_k^{\mathbf{X}} = \Sigma_{\mathbf{X}_k}^- \left(\mathbf{C}_k^{\mathbf{X}}\right)^T \left[\mathbf{C}_k^{\mathbf{X}} \Sigma_{\mathbf{X}_k}^- \left(\mathbf{C}_k^{\mathbf{X}}\right)^T + \Sigma_{\mathbf{v}_k}\right] \\ \hat{\mathbf{X}}_k = \hat{\mathbf{X}}_k^- + \mathbf{K}_k^{\mathbf{X}} \left[\mathbf{Y}_k - \mathbf{G}\left(\hat{\mathbf{X}}_k^-, \hat{\boldsymbol{\theta}}_k^-, \mathbf{u}_k\right)\right] \\ \Sigma_{\mathbf{X}_k} = \left(\mathbf{I} - \mathbf{K}_k^{\mathbf{X}} \mathbf{C}_k^{\mathbf{X}}\right) \Sigma_{\mathbf{X}_k}^- \end{cases}, \quad (16)$$

where $\mathbf{C}_k^{\mathbf{X}} = \dfrac{\partial \mathbf{G}(\mathbf{X}, \hat{\boldsymbol{\theta}}_k^-, \mathbf{u}_k)}{\partial \mathbf{X}}\bigg|_{\mathbf{X}=\hat{\mathbf{X}}_k^-}$.

Parameter update:

$$\begin{cases} \mathbf{K}_k^{\boldsymbol{\theta}} = \Sigma_{\boldsymbol{\theta}_k}^- \left(\mathbf{C}_k^{\boldsymbol{\theta}}\right)^T \left[\left(\mathbf{C}_k^{\boldsymbol{\theta}}\right) \Sigma_{\boldsymbol{\theta}_k}^- \left(\mathbf{C}_k^{\boldsymbol{\theta}}\right)^T + \Sigma_{\mathbf{v}_k}\right] \\ \hat{\boldsymbol{\theta}}_k = \hat{\boldsymbol{\theta}}_k^- + \mathbf{K}_k^{\boldsymbol{\theta}} \left[\mathbf{Y}_k - \mathbf{G}\left(\hat{\mathbf{X}}_k, \hat{\boldsymbol{\theta}}_k^-, \mathbf{u}_k\right)\right] \\ \Sigma_{\boldsymbol{\theta}_k} = \left(\mathbf{I} - \mathbf{K}_k^{\boldsymbol{\theta}} \mathbf{C}_k^{\boldsymbol{\theta}}\right) \Sigma_{\boldsymbol{\theta}_k}^- \end{cases}, \quad (17)$$

where $\mathbf{C}_k^{\boldsymbol{\theta}} = \dfrac{d\mathbf{G}(\hat{\mathbf{X}}_k, \boldsymbol{\theta}, \mathbf{u}_k)}{d\boldsymbol{\theta}}\bigg|_{\boldsymbol{\theta}=\hat{\boldsymbol{\theta}}_k^-}$.

### 3.2 The sequential algorithm

Incorporating the algorithms described in section 3.1, the sequential parameter and SoC/SoH estimation approach is formulated in 3 steps.

**Step #1:** The first step of the sequential algorithm is estimating the ohmic resistance by using a high-pass filter and injecting high-frequency current. Based on Eq. (6), the battery terminal voltage can be simplified as

$$V_{\text{bf}}(s) = -R_s i_{\text{bf}}(s). \quad (18)$$

Therefore, the discrete time state-space equation (9) for estimating the ohmic resistance $R_s$ using the EKF can be given as

$$\begin{cases} R_s(k) = R_s(k-1) + r_k \\ V_{\text{bf}}(k) = -R_s i_{\text{bf}}(k) + v_k \end{cases}. \quad (19)$$

**Step #2:** When the medium-frequency current is injected, the battery terminal voltage is governed by ohmic resistance dynamics and RC pair dynamics, so Eq. (6) can be simplified as

$$V_{bf}(s) = -R_s i_{bf}(s) - \frac{R_t}{1+\tau s} i_{bf}(s). \quad (20)$$

The estimated ohmic resistance obtained in Step #1 is then used in Step #2, and the bilinear transformation is used to discretize Eq. (20). Consequently, the discrete time state-space equation (11) for estimating $R_t$ and $\tau$ is given as

$$\begin{cases} \boldsymbol{\theta}_2(k) = \boldsymbol{\theta}_2(k-1) + \mathbf{r}_k \\ V_{bf}(k) = -\hat{R}_s(k) i_{bf}(k) - R_t i_2(k) + \mathbf{v}_k \end{cases}, \quad (21)$$

where

$$\begin{cases} \boldsymbol{\theta}_2(k) = \begin{bmatrix} R_t(k) & \tau(k) \end{bmatrix}^T \\ i_2(k) = \dfrac{T_s}{T_s + 2\tau} \left[ i_{bf}(k) + i_{bf}(k-1) \right] - \dfrac{T_s - 2\tau}{T_s + 2\tau} i_2(k-1) \end{cases},$$

and $T_s$ is the sampling period (i.e., 1s).

**Step #3:** Given that $R_s$, $R_t$, and $\tau$ are estimated in Steps #1 and #2, the SoC and SoH can be simultaneously estimated in Step #3. Based on Eqs. (1)-(3), the discrete time state-space equation for estimating the SoC and SoH estimation is shown as

$$\begin{cases} Q_b(k) = Q_b(k-1) + r_k \\ \mathbf{X}_3(k) = \begin{bmatrix} e^{-\frac{T_s}{\tau}} & 0 \\ 0 & 1 \end{bmatrix} \mathbf{X}_3(k-1) + \begin{bmatrix} \hat{R}_t \left( 1 - e^{-\frac{T_s}{\tau}} \right) \\ -\dfrac{\eta T_s}{Q_b} \end{bmatrix} i_b(k), \quad (22) \\ V_b(k) = \text{OCV}(z(k)) - v_C(k) - \hat{R}_s(k) i_b(k) \end{cases}$$

where

$$\mathbf{X}_3(k) = \begin{bmatrix} v_C(k) & z(k) \end{bmatrix}^T.$$

As shown in Eq. (22), the voltage of the RC pair $v_C$ is also estimated for better performance [2], and the DEKF presented in Section 3.1 is adopted. Since SoC and SoH are estimated together, an inaccurate estimation of either SoC or SoH will influence the other. However, the estimation accuracy can still be significantly improved for the proposed algorithm, since less uncertainties are involved in this case when compared to the case where all battery parameters/states are estimated simultaneously, as will be verified through experimental results later.

For different battery cells, which are manufactured by different companies or have different chemistries, the battery parameters may vary significantly. As a result, the frequency of the injected current and the cut-off frequency of the high-pass filter should be adapted to specific battery cells. At any rate, the proposed sequential algorithm is a general method for battery parameter/state estimation.

## 4. Experimental results

Experimental testing is focused on an 18650 Lithium ion battery (see Table 1 for detailed information). In the experiments, error caused by ECM inaccuracy is introduced. An Arbin battery test system (BT2000) was used in the following experiments [36]. The initial SoC is 80% for all three experiments, and they are illustrated as follows.

**Experiment #1:** Validation of the sequential algorithm under 20°C. In Experiment #1, the initial guesses of the estimated parameters are $[\hat{R}_s(0)\ \hat{R}_t(0)\ \hat{\tau}(0)\ \hat{Q}_b(0)]$=[0.02 0.03 15 2], and the initial values of estimated SoC and $v_C$ are 50% and 0V, respectively. As shown in Fig. 4(a), in Step #1 a 0.5Hz sinusoidal current with amplitude of 0.5C is injected and the first-order butterworth high-pass filter has 3dB bandwidth of 0.05Hz. The estimation result of $R_s$ is shown in Fig. 4(b), which indicates that the estimated value can accurately track the real value, which is obtained from the HPPC test.

In Step #2 a 0.02Hz sinusoidal current with amplitude of 0.5C is added onto a base current (0.004Hz sinusoidal current with amplitude of 0.5C) and the first-order butterworth high-pass filter has 3dB bandwidth of 0.002Hz. The estimated $R_s$ from Step #1 is used in Step #2, and the estimated parameters can track the actual values for $R_t$ and $\tau$, but there is a slight static error in the $R_t$ estimation result, as shown in Fig. 4(c). Therefore, the first-order ECM has limitations representing the studied 18650 Lithium ion battery, since the RC pair parameters are hard to estimate. Since the sampling frequency in Step #1 is high, we need to manually control the Arbin test system and switch to Step #2 in the experiment. As a result, there is a transient period from 200s to 287s, which is shown in Fig. 4(a). The estimation process of Step #2 is from 288s to 1187s, which includes a hold-up time of 400s to avoid initial SoC dynamics. Step #3 starts from 1200s.

The estimated $R_s$, $R_t$, and $\tau$ are used in Step #3 to estimate SoC and SoH simultaneously. A scaled New European Driving Cycle (NEDC) is used in Step #3 to represent a practical profile for batteries. As shown in Fig. 4(d), the SoC estimation performance is satisfactory and the estimated error is below 1% under a significant initial condition error (30%). As shown in Fig. 4(e), the estimated battery capacity accurately tracks the actual value obtained from the static capacity test after 1800s, and there is no significant error after convergence. The estimated voltage shown in Fig. 4(f) tracks the actual terminal value well. As a result, the effectiveness of the proposed sequential algorithm is verified experimentally.

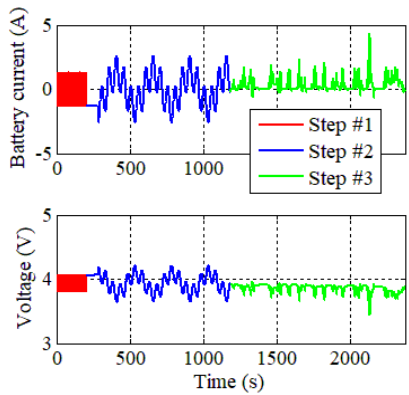

(a) The battery current profile

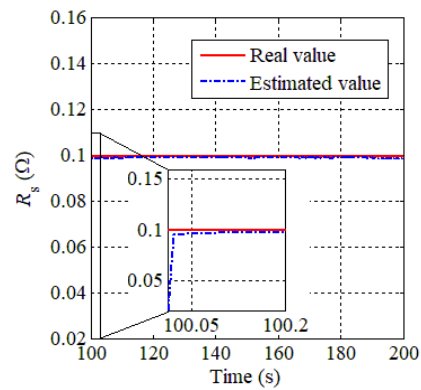

(b) $R_s$ estimation result (Step #1)

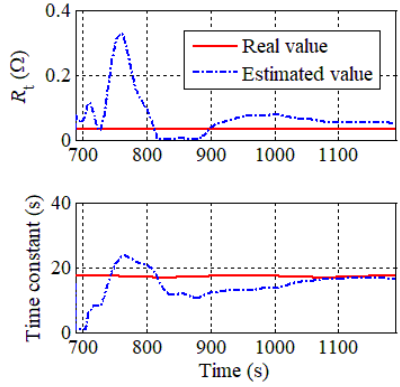
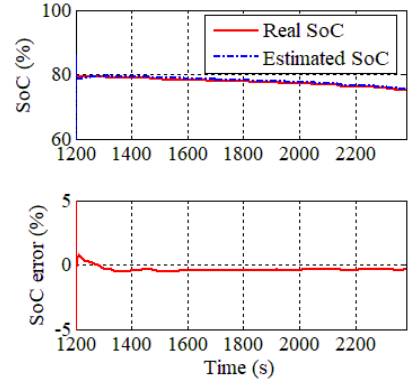

(c) $R_t$ and $\tau$ estimation result (Step #2)

(d) SoC estimation result (Step #3)

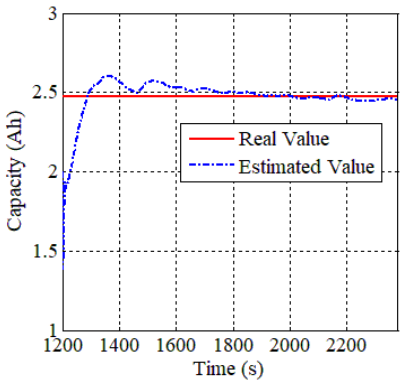
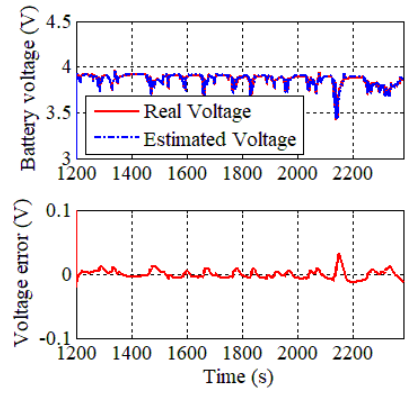

(e) $Q_b$ estimation result (Step #3)

(f) $v_b$ estimation result (Step #3)

Fig. 4. Experimental results Experimental results under 20°C

**Experiment #2:** Validation of the sequential algorithm under 40°C. In Experiment #2, the temperature is increased to 40°C, and the battery parameters slightly change when compared to the those of 20°C according to the battery static capacity test [37]. Especially, the battery capacity increases from 2.47Ah to 2.62Ah. The initial guesses of the estimated parameters and states are similar to the ones in Experiment #1.

The same current profile, as shown in Fig. 4(a), as well as the same high-pass filters are used. As shown in Fig. 5(a) and (b), the parameter estimation performance is satisfactory and similar to the results of Experiment #1. The SoC estimation error is less than 1% when the initial guess error is 30%, as shown in Fig. 5(c). For the battery capacity (i.e., SoH), the convergence time is about 600s, and no significant static error exists after convergence, as shown in Fig. 5(d).

The proposed algorithm not only decreases the computational cost when compared to the case where all parameter/states are estimated simultaneously, but also improves the estimation performance. This conclusion can be theoretically proven using Cramer-Rao (CR) bound analysis [38]. It has been shown that the estimation error is increased when more parameters are considered in the estimation [39]. Therefore, a single-parameter estimation is the most accurate as the least amount of uncertainties are involved in the estimation process. The proposed sequential algorithm can significantly improve the estimation accuracy when compared to the multi-parameter estimation, given that the ECM dynamics dominated by different components can be separated [4]. Experiment #3 is conducted to verify the proposed sequential algorithm.

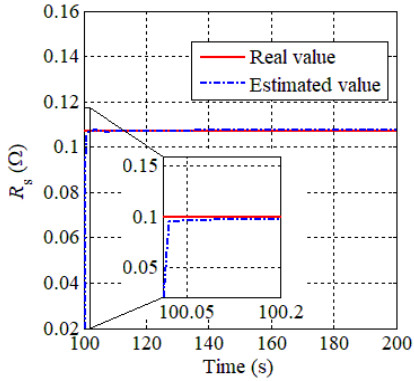
(a) $R_s$ estimation result (Step #1)

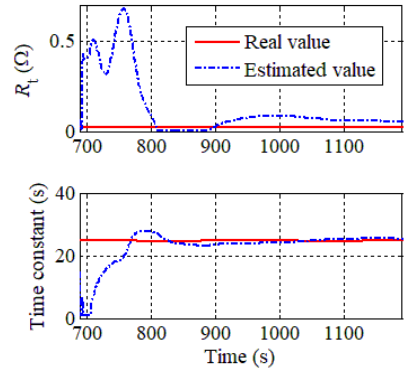
(b) $R_t$ and $\tau$ estimation result (Step #2)

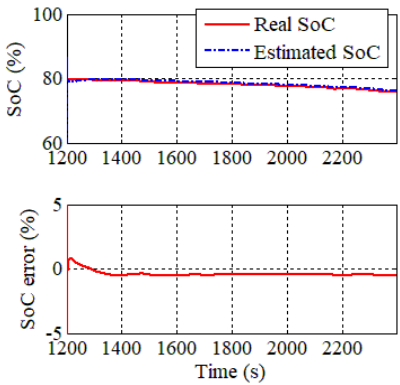
(c) SoC estimation result (Step #3)

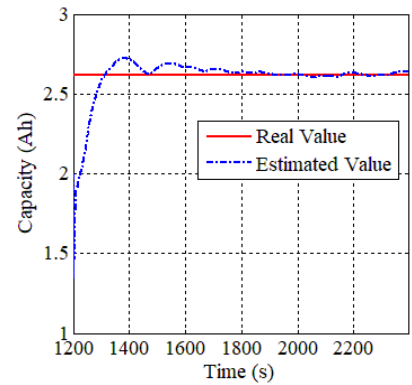
(d) $Q_b$ estimation result (Step #3)

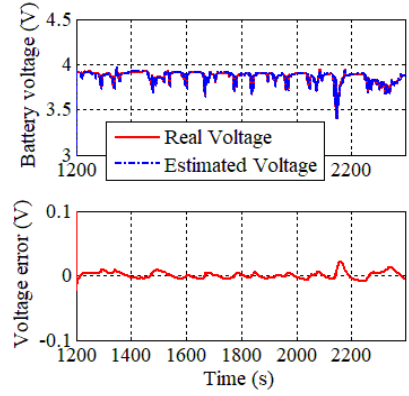

(e) $v_b$ estimation result (Step #3)

Fig. 5. Experimental results under 40°C

**Experiment #3:** Performance of concurrent parameter/state estimation under 20°C. The multi-scale EKF is used to estimate all parameters/states simultaneously [2], [34]. When compared to the DEKF, the multi-scale EKF estimates the parameters much slower than it estimates the states because the parameters generally vary slowly [34]. The optimal current profile consisting of three sine waves (i.e., 0.01Hz, 0.05Hz, and 0.1Hz) is used in Experiment #3, as shown in Fig. 6(a). Detailed information on determining these optimal current frequencies is provided in [39]. The parameter estimation results are shown in Fig. 6(b), revealing that the estimation performance of $R_s$ and $Q_b$ (i.e., SoH) are satisfactory. The estimated $R_t$ and $\tau$ cannot track the actual values even using the optimal data, which can theoretically achieve the best estimation performance. As shown in Fig. 6(c), the estimated SoC needs a longer time to converge to the real value as compared to the sequential algorithm. Moreover, the static error of the estimated SoC is around 2%. The estimation error of the battery terminal voltage is correspondingly enlarged when more parameters are estimated, as shown in Figs. 4(f), 5(e), and 6(d). Based on the above experimental results, it is shown that the sequential algorithm, which separates the estimation process, can achieve a better estimation performance when compared to the case where all parameters/states are estimated simultaneously.

We would like to point out that the implications of injecting a current signal for active parameter estimation could be complicated for general application (e.g., electric vehicles). However, an over-actuated system like the battery/supercapacitor (SC) hybrid energy storage system [40] or hybrid electric vehicle [41] provides an opportunity to inject desired signals for identification and achieve control objectives simultaneously [42]. Therefore, the proposed sequential algorithm can be directly used, and the potential negative influence of injecting the current on the system performance (i.e., system efficiency and power supply quality) can be minimized given the over-actuated nature. Specifically, for any power demand $P_d$, we have $P_d = P_{s1} + P_{s2}$, where $P_{s1}$ and $P_{s2}$ denote the power from source #1 (i.e., battery) and source #2 (i.e., SC). The required current for battery parameter/state estimation can be injected directly, while the SC can compensate to ensure the entire system supplies the demanded power. In addition, when the battery is used as the sole energy source, the proposed algorithm also can be used if the battery charging current can be changed and therefore the required excitation can be added. The influence of temperature is investigated in the experiment, and the remarks are given to address the influence of battery degradation.

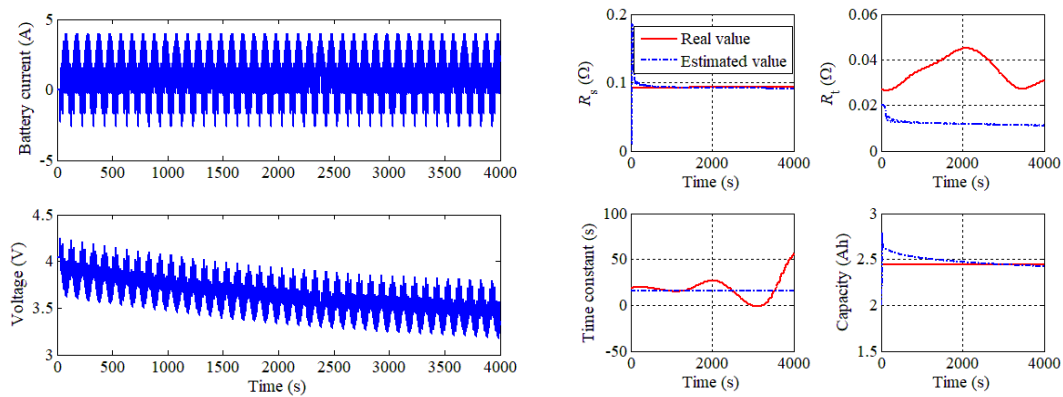

(a) Optimal current profile  (b) Parameters estimation result

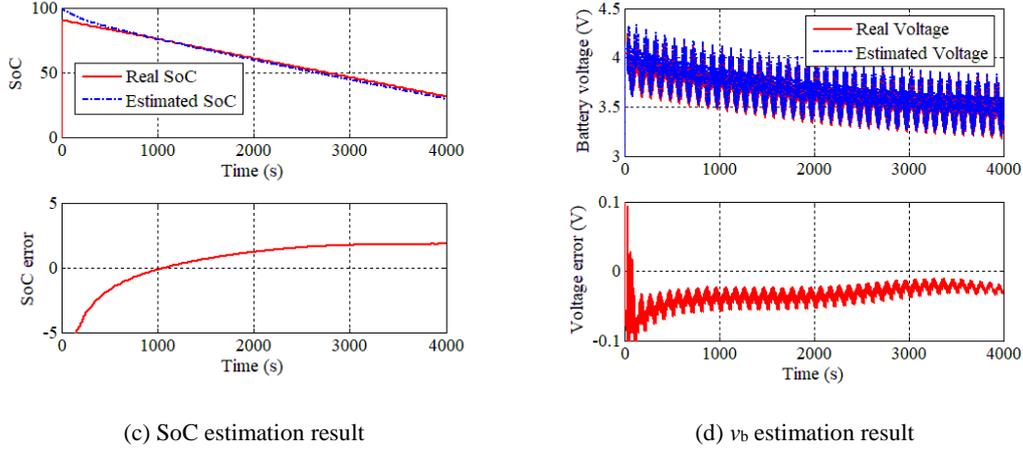

(c) SoC estimation result      (d) $v_b$ estimation result

Fig. 6. Experimental results of estimating all parameters/states simultaneously

**Remark 1.** As a battery ages, its parameters will change, and the change will be reflected in the estimation results. One of the main goals for online parameter estimation is to detect aging for condition monitoring. It has been proven that the proposed sequential algorithm can accurately estimate the battery parameters when they change due to temperature and SoC variations. Therefore, the proposed algorithm can effectively detect the battery degradation in practical applications.

Another remark is given below to highlight the novelty of the proposed sequential algorithm when compared to existing methods.

**Remark 2.** As mentioned above, battery parameters vary with working conditions (e.g., temperature) and battery degradation levels. The results provided in [43] show promising experimental results for battery SoC and SoH estimation when the parameters can be calibrated offline and used online. However, it is challenging to calibrate battery parameters for all conditions (e.g., different temperatures and degradation levels). So the improved method is still required for practical applications. When the battery is adopted in over-actuated systems, the proposed sequential algorithm can solve aforementioned problem well by actively injecting persistently exciting signals, as shown in Figs. 4 and 5. It means that massive calibration work is avoided by adopting the proposed algorithm, while the SoC/SoH estimation performance can still be ensured.

## 5. Conclusion

When battery states/parameters are estimated simultaneously, substantial uncertainties are introduced in the estimation process, and inaccurate parameters can therefore impair the state estimation performance. To improve estimation performance, the sequential algorithm, which uses frequency-scale separation and estimates the parameters/states sequentially by injecting the current with different frequencies, is proposed in this paper. Specifically, by using a high-pass filter, the ohmic resistance $R_s$ can be estimated independently via injecting a high-frequency current. Then, using the estimated $R_s$, the RC pair can be estimated by injecting a medium-frequency current. Finally, based on the above estimated parameters, the battery SoC and SoH can be estimated simultaneously. Experimental results show that the estimation accuracy of the proposed sequential algorithm is satisfactory and better than the case where all parameters/states are estimated simultaneously. The proposed algorithm can be implemented online when the battery is used in over-actuated systems.

## Acknowledgement

This work is supported by the National Natural Science Foundation of China (Grant No. 51877057).